\documentclass[aps,pra,showpacs,secnumarabic]{revtex4}

\usepackage{amsmath}
\usepackage{amssymb}
\usepackage[cp1251]{inputenc} 
\usepackage{bm}
\usepackage{graphicx}
\def\ZETF{Zh. Eksp. Teor. Fiz.\ }
\def\SPJ{Sov. Phys.-JETP\ }
\def\JPCM{J. Phys.: Condens. Matter\ }

\newcommand{\bra}[1]{\ensuremath{\bm{\langle}#1\bm{|}}}
\newcommand{\ket}[1]{\ensuremath{\bm{|}#1\bm{\rangle}}}

\begin{document}

\title{One-particle model of the metal-insulator transition\\
       in two-dimensional systems at $T=0$}
\title{The metal-insulator transition in 2D systems at $T=0$:\\
one-particle approach}

\author{Yu.\,V.\,Tarasov}
\email{yutarasov@ire.kharkov.ua}

\affiliation{Institute for Radiophysics \& Electronics NAS of Ukraine,\\
12 Acad. Proskura St., 61085 Kharkov, Ukraine}

\date{\today}

\begin{abstract}
The conductance of a disordered finite-size electron system is
calculated by reducing the initial dynamic problem of arbitrary
dimensionality to strictly one-dimensional problems for
one-particle mode propagators. The metallic ground state of a
two-dimensional conductor, which is considered as a limiting case
of the actually three-dimensional quantum waveguide, is shown to
result from its multi-modeness. On lowering the waveguide
thickness, in practice, e.g., due to application of the
``pressing'' potential (depletion voltage), the electron system
undergoes a set of continuous phase transitions connected with the
discrete change in the number of extended modes. The closing of
the last current-carrying mode is interpreted as the electron
system transition from metallic to dielectric state. The results
obtained agree qualitatively with the observed ``anomalies'' of
the resistance of different electron and hole systems.
\end{abstract}



\maketitle
\section{INTRODUCTION}

The problems associated with the electron transport in disordered
systems have for years been highly attracting for many
researchers. This is because they are crucially important from the
applicability standpoint and, besides, they bring forward
intriguingly challenging tasks arising in this field. One of these
problems, which has not been unambiguously solved until the
present time, applies to the nature of the unusual phenomenon
observed in two-dimensional electron and hole systems which is by
many authors interpreted as a metal-insulator transition (MIT)
associated with the disorder and the interaction of current
carriers. The unusual properties of the conductance of different
planar hetero-structures (see extensive bibliography in
\cite{bib:AKS01}) are clearly at variance with a common conviction
that there can be no metallic ground state in two-dimensional (2D)
systems, in the same way as in one-dimensional (1D) ones, at any
strength of disorder \cite{bib:AALR79}.

Numerous attempts aiming to explain the ``anomalous''
low-temperature metallic behaviour of 2D electron and hole systems
were made using different physical ideas. Among these were the
existence of a conducting state of heavily dilute electrons
\cite{bib:F84,bib:CCL98}, their non-Fermi-liquid behaviour
\cite{bib:CYA98}, the possibility of a superconductive state of 2D
interacting electrons \cite{bib:BK98,bib:TN98},
temperature-dependent screening of the electron-impurity
scattering \cite{bib:KS99,bib:SH00}, etc. However, the fundamental
question of whether the observed resistance anomalies should be
considered exhibiting a \emph{true quantum phase
transition}~\cite{bib:SGCS97} or they can be explained within a
framework of the conventional theory of disordered
systems~\cite{bib:LR85} remains open.

In this paper, the model for describing the observed phenomena is
proposed which actually realizes the concept of electron states
quantum dephasing due to the interaction of those electrons with
some ``dephasing environment'' whose intrinsic state is not traced
in the course of the experiment \cite{bib:MOH99}. It is generally
believed that the loss of electron coherence in conductors subject
to quenched disorder is always caused by conventional inelastic
scattering processes (electron-phonon, electron-electron, etc.).
As a result, the corresponding dephasing rates vanish when the
temperature approaches zero. However, from recent publications it
has become clear that the physical nature of dephasing environment
for real systems still remains controversial \cite{bib:AAG99}.
Some authors regard the quasi-elastic Coulomb interaction of
carriers as the most probable cause of dephasing of the initially
coherent (presumably localized) electron states, since the
``anomalous'' behaviour of the resistance is commonly observed in
2D systems of low electron density ($r_s\gtrsim 10$,
$r_s=E_{e-e}/E_F$ is the ratio of Coulomb energy to Fermi energy
of the electrons). However, this kind of interaction is quite
differently evaluated by different theories, viz. some authors
reckon it as promoting localization \cite{bib:AAL80,bib:TC89}
whereas some as inhibiting its origination
\cite{bib:F84,bib:CCL98,bib:BK94}.

Meanwhile, it was shown in \cite{bib:Tar99,bib:Tar00} that
scattering from quenched disorder can lead , in much the same way
as inelastic processes do, to the dephasing of quantum states
properly classified with regard for the confinement of the real
dynamical system being considered. The evidence was based on the
use of the mode representation for one-particle propagators, which
seems to be most appropriate as applied to \emph{open systems} of
waveguide configuration. Of no less significance with regard to
electrons in solids is the fact that the mode states represent
\emph{collective} excitations which are well adapted for
describing a system of strongly correlated current carriers. As a
matter of fact, the electron correlation, even without invoking
Coulomb interaction, is originally embedded in the theory if one
applies the Green function formalism which explicitly takes into
account the Pauli principle \cite{bib:MA00}.

In Refs.~\cite{bib:Tar99,bib:Tar00}, it was shown that in
not-too-narrow 2D conductors, when there is the availability of
more than one \emph{extended} mode (or, in other words, more than
one conducting channel), the inter-mode scattering, unless it is
suppressed by virtue of some peculiar circumstances, leads to the
dephasing of coherent mode states, thus preventing them from
interference localization. In this case, a set of open channels
other than the selected one serves as a dephasing environment. In
the case where the inter-mode scattering is non-existent, which is
valid, e.g., for conductors that are randomly layered in the
direction of current, the electron localization (of Anderson type)
occurs in each of the channels independently. This results in the
exponential fall (well-known from quasi-one-dimensional systems
theory \cite{bib:Efet83,bib:D84,bib:MPK88}) of the conductance
against the conductor length when the latter exceeds the value of
the order of $N_c\ell$, with $N_c$ being the number of open
channels and $\ell$ the electron mean-free path.

Although from the results given in \cite{bib:Tar99,bib:Tar00} it
follows that the metallic ground state of two-dimensional
quench-disordered systems should not be considered to be an
anomalous phenomenon, the physics of a 2D system transition from
conducting to dielectric state, which is observed in numerous
experiments, was not identified. In this paper, to ascertain the
physical nature of MIT in planar heterostructures it is suggested
to fit the formal statement of the problem to experimental
conditions by extending the method developed in
\cite{bib:Tar99,bib:Tar00} for exactly two-dimensional systems to
systems of higher dimensionality. Such an approach is motivated by
the fact that in practice 2D electron systems are mostly created
by forming a near-surface \emph{finite width} potential wells. The
well is normally produced either by means of the ``pressing''
external electric field or due to the contact potential.

\section{STATEMENT OF THE PROBLEM AND MODELLING THE 2D CONDUCTING SYSTEM}

\begin{figure}[h]
\centering \scalebox{.8}[.8]{\includegraphics{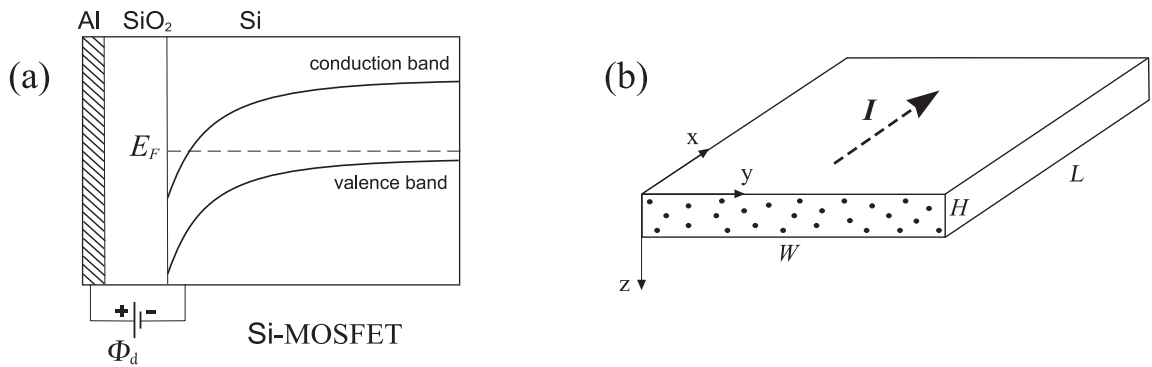}}
\caption{The actual (a) and the model (b) configuration of a 2D
conducting system} \label{Fig1}
\end{figure}

The conductors of reduced dimension (one- and two-dimensional)
provide a mathematical idealization of genuine physical objects
which are in fact geometrically three-dimensional. The potential
well resulting from the band bending in the region of different
materials contact (see, e.g., Fig.~\ref{Fig1}a) generate a
near-surface quantum waveguide of finite thickness, in which the
in-plane density of the carriers is normally varied either by
application of the external depletion voltage $\Phi_d$ or through
a capacitive control. The shape of the near-surface well (in most
cases it is nearly triangular~\cite{bib:BK94,bib:SK99})
is of no crucial importance for its principal mission, viz. to
confine the electrons along the direction perpendicular to the
heterophase boundary. Therefore, to simplify calculations, we
hereinafter consider a planar conductor in the form of a
rectangular three-dimensional ``electron waveguide'' with hard
side boundaries (Fig.~\ref{Fig1}b) which envelopes the region
\begin{equation}\label{geom}
  x\in(-L/2,L/2)\ ,\qquad y\in[-W/2,W/2]\ ,\qquad z\in[-H/2,H/2]
  \ .
\end{equation}
The length $L$, the width $W$, and the height $H$ of the waveguide
will be considered as arbitrary.

According to the linear response theory \cite{bib:K57}, the
dimensionless (in units of $e^2/\pi\hbar$) static conductance
$g(L)$ is expressed at $T=0$ in terms of one-particle electron
propagators in the following manner,
\begin{equation}\label{Kubo_main}
  g(L)=\frac{2}{L^2}\iint \mathrm{d}\mathbf{r}\,\mathrm{d}\mathbf{r}'
  \frac{\partial}{\partial x}\left[G^A(\mathbf{r},\mathbf{r}')-
  G^R(\mathbf{r},\mathbf{r}')\right]
  \frac{\partial}{\partial x'}\left[G^A(\mathbf{r}',\mathbf{r})-
  G^R(\mathbf{r}',\mathbf{r})\right] \ .
\end{equation}
Here, $G^{R,A}(\mathbf{r},\mathbf{r}')$ are the retarded (R) and
advanced (A) electron Green functions, the integration is taken
over the volume occupied by the conductor. Within the isotropic
Fermi liquid model and with the units such that $\hbar\hm=2m\hm=1$
($m$ is the electron effective mass) the retarded propagator,
whose index ``R'' will be henceforth omitted, is governed by the
equation
\begin{equation}\label{master}
  \left[
    \Delta+k_F^2+{\rm i}0-V(\mathbf{r})
  \right]
  G(\mathbf{r},\mathbf{r}')=\delta(\mathbf{r}-\mathbf{r}') \ .
\end{equation}
Here $\Delta$ is the three-dimensional Laplacian, $k_F$ is the
Fermi wavenumber, $V(\mathbf{r})$ is the random static potential
specified by the zero mean value, $\langle
V(\mathbf{r})\rangle=0$, and the binary correlation function
$\left<V(\mathbf{r})V(\mathbf{r}')\right>=
\mathcal{QW}(\mathbf{r}-\mathbf{r}')$. We assume the function
$\mathcal{W}(\mathbf{r})$ to be normalized to unity and decaying
at a characteristic scale $r_c$ (the correlation radius). In what
follows, for the sake of simplicity, we restrict our consideration
to the correlation function of less general form, namely
\begin{equation}\label{corr3d_simpl}
  \left<V(\mathbf{r})V(\mathbf{r}')\right>=\mathcal{QW}(x-x')
  \delta(\mathbf{r}_{\perp}-\mathbf{r}'_{\perp}) \ ,
  \qquad\mathbf{r}_{\perp}=(y,z)\ ,
\end{equation}
which evidently should not considerably affect the result.

The equation (\ref{master}) must be supplemented with proper
boundary conditions. The side boundaries of the system,
``impenetrable'' for electrons, can be characterized by a real
impedance which particularly corresponds to the Dirichlet
conditions
\begin{equation}\label{Dirichlet}
  G(\mathbf{r},\mathbf{r}')\Bigg|_{y=\pm W/2\atop z=\pm H/2}=0\ .
\end{equation}
At the same time, the conductor, once attached at the points
$x=\pm L/2$ to equilibrium ``reservoirs'', should be considered as
an open system, which implies two important consequences. First,
in Kubo theory the chemical potentials of the massive leads are
assumed to be equal to one another. That is why the chemical
potential of the junction connecting the leads (or, in the
conducting phase, the electron Fermi energy in that junction) can
be thought of as being independent of the waveguide geometry, thus
allowing to put hereinafter $k_F=const$. Second, the openness of
the waveguide butt ends imposes the condition for the contact
surface impedance to be complex-valued. This results in the
differential operation in (\ref{master}) being non-Hermitian.

In \cite{bib:Tar99,bib:Tar00}, a method was proposed for solving
such a non-Hermitian problem in two dimensions. The analogous
procedure is applicable to waveguide-type systems of arbitrary
dimensionality as well. The important element of the method is a
transition from one originally multi-dimensional stochastic
problem to an infinite set of strictly one-dimensional (in general
case, non-Hermitian) problems for the propagator
$G(\mathbf{r},\mathbf{r}')$ mode components. In the next section,
the main points of the method \cite{bib:Tar99,bib:Tar00} are set
forth as applied to the system being considered.

\section{REDUCTION TO ONE DIMENSION}
\subsection{The general scheme}

The suggested algorithm of the multi-dimensional problem
(\ref{master}) reduction to a set of exactly one-dimensional
boundary problems is applicable to open systems of arbitrary
waveguide configuration at any strength of disorder. The first
step consists in proceeding to the mode representation of the
electron propagators. In the case of the waveguide depicted in
Fig.~\ref{Fig1}b the transition is carried out by expanding all
functions in a whole set of the transverse Laplace operator
eigenfunctions, $\ket{\mathbf{r}_{\perp};\bm{\mu}}$, which are
composed of ordinary trigonometric functions. Assuming the
boundary conditions (\ref{Dirichlet}), these eigenfunctions can be
chosen in the form
\begin{equation}\label{transv_set}
  \ket{\mathbf{r}_{\perp};\bm{\mu}}=\frac{2}{\sqrt{WH}}
  \sin\left[\left(\frac{y}{W}+\frac{1}{2}\right)\pi n\right]
  \sin\left[\left(\frac{z}{H}+\frac{1}{2}\right)\pi m\right] \ ,
\end{equation}
where $\bm{\mu}=(n,m)$ is the vectorial mode index with the
components $n,m\in\aleph$. By the functions (\ref{transv_set}),
the equation (\ref{master}) is transformed into a set of coupled
equations for mode Fourier components of the function
$G(\mathbf{r},\mathbf{r}')$,
\begin{equation}\label{mode_eqn}
  \left[\frac{\partial^2}{\partial x^2}+
  \kappa^2_{\bm{\mu}}+{\rm i}0-
  V_{\bm{\mu}}(x)\right]G_{\bm{\mu}\bm{\mu}'}(x,x')
  -\sum_{\bm{\nu}\neq\bm{\mu}}U_{\bm{\mu}\bm{\nu}}(x)
  G_{\bm{\nu}\bm{\mu}'}(x,x')=
  \delta_{\bm{\mu}\bm{\mu}'}\delta(x-x') \ .
\end{equation}
In (\ref{mode_eqn}), the parameter
\begin{equation}\label{kappa_munu}
  \kappa^2_{\bm{\mu}}=k_F^2-\left({\pi n}/{W}\right)^2-
  \left({\pi m}/{H}\right)^2
\end{equation}
has the meaning of an unperturbed \emph{longitudinal energy} of
the mode $\bm{\mu}$. The potential matrix
$\|U_{\bm{\mu}\bm{\mu}'}\|$ is composed of the functions
\begin{equation}\label{Umatr}
  U_{\bm{\mu}\bm{\mu}'}(x)=\int_S\textrm{d}\mathbf{r}_{\perp}
  \ket{\mathbf{r}_{\perp};\bm{\mu}}
  V(\mathbf{r})\bra{\mathbf{r}_{\perp};\bm{\mu}'}
\end{equation}
where integration is taken over the conductor cross-section $S$.
The diagonal components of this matrix,
$V_{\bm{\mu}}(x)\equiv~U_{\bm{\mu}\bm{\mu}}(x)$, are responsible
for the \emph{intra-mode} whereas off-diagonal for the
\emph{inter-mode} scattering of quantum particles. In
Eq.~(\ref{mode_eqn}), the term containing the intra-mode potential
is deliberately separated from the terms with inter-mode
potentials $U_{\bm{\mu}\bm{\nu}}(x)$ ($\bm{\nu}\neq\bm{\mu}$) to
avoid singularities in the course of developing the perturbation
theory (the proof is presented in Ref.~\cite{bib:Tar00}).

The initial problem reformulated in terms of the one-coordinate
differential equations (\ref{mode_eqn}) cannot actually be
considered as strictly one-dimensional due to the entanglement of
all mode components of the Green function
matrix~$\|G_{\bm{\nu}\bm{\mu}}\|$. To reduce
equation~(\ref{master}) to a set of \emph{independent}
one-dimensional equations introduce, in the first stage, the
auxiliary mode propagators allowing for the scattering by
intra-mode potentials only,
\begin{equation}\label{trialGV}
  \left[\frac{\partial^2}{\partial x^2}+
  \kappa^2_{\bm{\mu}}+{\rm i}0-
  V_{\bm{\mu}}(x)\right]G^{(V)}_{\bm{\mu}}(x,x')
  =\delta(x-x') \ .
\end{equation}
For ``trial'' Green function $G^{(V)}_{\bm{\mu}}(x,x')$, the
demand of the waveguide openness at the ends $x=\pm L/2$ can be
formulated in terms of Sommerfeld's radiation conditions
\cite{bib:BF,bib:V67}. Assuming the contact between the conductor
and leads to be ideal (not resulting in scattering), these
conditions appear in the form
\begin{equation}\label{rad_cond}
  \left(\frac{\partial}{\partial x}\mp {\rm
  i}\kappa_{\bm{\mu}}\right)
  G^{(V)}_{\bm{\mu}}(x,x')\Bigg|_{x=\pm L/2}=0 \ ,
  \qquad\qquad x'\in(-L/2,L/2)\ .
\end{equation}

Regarding hereinafter the solution of (\ref{trialGV}),
(\ref{rad_cond}) as known precisely, it is worth passing from the
differential equation~(\ref{mode_eqn}) to the integral equation,
\begin{equation}\label{Gmatr_eq}
G_{\bm{\mu}\bm{\mu}'}(x,x')=
G^{(V)}_{\bm{\mu}}(x,x')\delta_{\bm{\mu}\bm{\mu}'}
+\sum_{\bm{\nu}\neq\bm{\mu}}\int_L\textrm{d}x_1\, \mathsf
R_{\bm{\mu}\bm{\nu}}(x,x_1) G_{\bm{\nu}\bm{\mu}'}(x_1,x') \ ,
\end{equation}
whose kernel,
\begin{equation}\label{kernR}
  \mathsf
  R_{\bm{\mu}\bm{\nu}}(x,x')=G^{(V)}_{\bm{\mu}}(x,x')U_{\bm{\mu}\bm{\nu}}(x')
  \ ,
\end{equation}
contains the potential $V(\mathbf{r})$ inter-mode harmonics only.
From Eq.~(\ref{Gmatr_eq}), all of the matrix
$\|G_{\bm{\nu}\bm{\mu}}\|$ off-diagonal elements can be expressed
in terms of the corresponding diagonal elements by means of some
linear operator $\hat{\mathsf K}$ specified in coordinate-mode
space $\mathsf{M}=\{x,\bm{\nu}\}$,
\begin{equation}\label{G_nm->G_mm}
  G_{\bm{\nu}\bm{\mu}}(x,x')=\int_L\mathrm{d}x_1\,
  \mathsf K_{\bm{\nu}\bm{\mu}}(x,x_1)G_{\bm{\mu}\bm{\mu}}(x_1,x') \
  ,\qquad\qquad\bm{\nu}\neq\bm{\mu} \ .
\end{equation}
The matrix elements of that operator, $\mathsf
K_{\bm{\nu}\bm{\mu}}(x,x')$, satisfy the multi-channel
Lippmann-Schwinger equation~\cite{bib:T75}
\begin{equation}\label{K_numu}
  \mathsf K_{\bm{\nu}\bm{\mu}}(x,x')={\mathsf R}_{\bm{\nu}\bm{\mu}}(x,x')
  +\sum_{\bm{\nu}_1\neq \bm{\mu}}
  \int_Ldx_1\,{\mathsf R}_{\bm{\nu}\bm{\nu}_1}(x,x_1)
  {\mathsf K}_{\bm{\nu}_1\bm{\mu}}(x_1,x') \ ,
\end{equation}
whose solution in operator form, $\hat{\mathsf
K}=(\openone-\hat{\mathsf R})^{-1}\hat{\mathsf R}$, is expressed
in terms of the operator $\hat{\mathsf R}$ given in $\mathsf{M}$
by the matrix elements~(\ref{kernR}).

Note that the lack of the terms with mode index $\bm {\mu}$ in
sums of (\ref{Gmatr_eq}) and (\ref{K_numu}) permits interpreting
the operator $\hat{\mathsf R}$ as an inter-mode scattering
operator acting in the reduced coordinate-mode space
${\mathsf{\overline M}_{\bm{\mu}}}$ that includes all the quantum
waveguide modes except the mode $\bm{\mu}$. The presence of mode
index $\bm{\mu}$ in the kernel of integral operator
(\ref{G_nm->G_mm}), as well as in other appropriate positions,
will be ensured by the projection operator $\bm{P}_{\bm{\mu}}$
that will make the mode index of any operator standing next to it
(both from the left and right) equal to the given value $\bm
{\mu}$.

By putting $\bm{\mu}'=\bm{\mu}$ in Eq.~(\ref{mode_eqn}) and
substituting the inter-mode propagators in the form
(\ref{G_nm->G_mm}) we eventually arrive at a closed
\emph{one-dimensional} differential equation for the diagonal
propagator $G_{\bm{\mu}\bm{\mu}}(x,x')$,
\begin{equation}\label{GDIAG-FIN}
  \left[\frac{\partial^2}{\partial x^2}+
  \kappa^2_{\bm{\mu}}+{\rm i}0-V_{\bm{\mu}}(x)-\hat{\mathcal T}_{\bm{\mu}}\right]
  G_{\bm{\mu}\bm{\mu}}(x,x')=\delta(x-x') \ .
\end{equation}
Here, along with the ``prime'' intra-mode potential
$V_{\bm{\mu}}(x)$, the effective \emph{non-local} (operator in
$x$-space) potential has arisen,
\begin{equation}\label{T-oper}
  \hat{\mathcal T}_{\bm{\mu}}=\bm{P}_{\bm{\mu}}\hat{\mathcal U}
  (\openone-\hat{\mathsf R})^{-1}\hat{\mathsf R}\bm{P}_{\bm{\mu}}=
  \bm{P}_{\bm{\mu}}\hat{\mathcal U}
  (\openone-\hat{\mathsf R})^{-1}\bm{P}_{\bm{\mu}} \ ,
\end{equation}
where $\hat{\mathcal U}$ stands for the inter-mode operator
potential specified in ${\mathsf{\overline M}_{\bm{\mu}}}$ by the
matrix elements
\begin{equation}\label{operU}
  \ket{x,\bm{\mu}}\hat{\mathcal U}\bra{x',\bm{\nu}}=
  U_{\bm{\mu}\bm{\nu}}(x)\delta(x-x') \ .
\end{equation}

Strictly speaking, the potential $\hat{\mathcal T}_{\bm{\mu}}$,
just like $V_{\bm{\mu}}(x)$, is the intra-mode one in that both
the initial and the final scattering states belong to the mode
$\bm{\mu}$. At the same time, this potential takes exactly into
account the \emph{inter-mode} scattering. From the
operator~(\ref{T-oper}) structure it can be seen that the
scattering caused by $\hat{\mathcal T}_{\bm{\mu}}$ can be regarded
as occurring through the intermediate ``trial'' mode states
corresponding to propagators $G^{(V)}_{\bm{\nu}}(x,x')$ with
$\bm{\nu}\neq\bm{\mu}$. Therefore, the operator $\hat{\mathcal
T}_{\bm{\mu}}$ will be termed henceforth the \emph{inter-mode}
potential. From the mathematical point of view it is nothing but
the conventional \textit{T}-matrix well known from the quanum
theory of scattering~\cite{bib:N68,bib:T75}.

At the final stage of the reduction of the multi-dimensional
conductance problem to the one-dimensional problem
(\ref{GDIAG-FIN}) it is worth representing expression
(\ref{Kubo_main}) directly through the functions
$G_{\bm{\mu}\bm{\mu}}(x,x')$. By expanding the electron
propagators in terms of eigenfunctions (\ref{transv_set}) one can
discriminate two different terms of the conductance. Within the
first term, henceforth conditionally called the ``diagonal''
conductance, $g^{(d)}(L)$, we collect those expansion terms which
from the outset contain the diagonal mode propagators
$G_{\bm{\mu}\bm{\mu}}$. All the other expansion terms, containing
mode components $G_{\bm{\nu}\bm{\mu}}$ with
$\bm{\nu}\neq\bm{\mu}$, will be gathered in the second term, the
``non-diagonal'' conductance $g^{(nd)}(L)$. Taking into account
the relationship (\ref{G_nm->G_mm}) and the fact that both the
retarded and advanced Green functions of the evanescent modes
($\kappa^2_{\bm{\mu}}<0$) at weak scattering can be regarded as
real-valued (see Eq.~(\ref{evan}) in the next subsection) the
above-indicated terms of the conductance can be represented as
\begin{subequations}\label{D-ND_conds}
\begin{eqnarray}\label{D_cond}
  &&g^{(d)}(L)=
  -\frac{4}{L^2}\overline{\sum_{\bm{\mu}}}\iint_L
  \mathrm{d}x\,\mathrm{d}x' \frac{\partial
  G_{\bm{\mu}\bm{\mu}}(x,x')}{\partial x} \frac{\partial
  G_{\bm{\mu}\bm{\mu}}^*(x,x')}{\partial x'}\ ,\\
  \label{nD_cond}
  &&g^{(nd)}(L) =-\frac{4}{L^2}\overline{\sum_{\bm{\mu},\bm{\nu}
  \atop\bm{\nu}\neq\bm{\mu}}}\iiiint_L
  \mathrm{d}x_1\ldots\mathrm{d}x_4\, \frac{\partial {\sf
  K}_{\bm{\nu}\bm{\mu}}(x_1,x_2)}{\partial
  x_1}G_{\bm{\mu}\bm{\mu}}(x_2,x_4) {\sf
  K}_{\bm{\nu}\bm{\mu}}^*(x_1,x_3)\frac{\partial
  G_{\bm{\mu}\bm{\mu}}^*(x_3,x_4)}{\partial x_4}\ .
\end{eqnarray}
\end{subequations}
The bar over the sum symbols in (\ref{D-ND_conds}) indicates the
summation over \emph{extended} modes only, i.e. the modes with
mode energies~$\kappa^2_{\bm{\mu}}>0$.

\subsection{The weak scattering approximation}

In view of statistical formulation of the problem, the important
ingredient of the calculation technique, viz., the trial Green
function $G^{(V)}_{\bm{\mu}}(x,x')$, can be thought of as known
precisely if one manages to find all its statistical moments
$\langle\big[G^{(V)}_{\bm{\mu}}(x,x')\big]^{p}\rangle$,
$p\in\aleph$. In the case of a strongly disordered system this can
certainly be done with the aid of numerical methods only. Yet
provided the scattering from the potential $V(\mathbf{r})$ is
regarded as weak, these moments can be obtained analytically
using, e.g., the method of \cite{bib:Tar00} which takes properly
into account the multiple scattering in the stochastic problem
(\ref{trialGV}), (\ref{rad_cond}).

The weakness criteria can be formulated in terms of the
inequalities
\begin{equation}\label{weakness}
  k_F,\,r_c\ll\ell\ ,
\end{equation}
where $\ell$ stands for the quasi-classical mean free path of
conducting electrons. In the particular case of a white-noise-type
potential, i.e. at $\mathcal{W}(x)=\delta(x)$ in
(\ref{corr3d_simpl}), this length equals to $4\pi/\mathcal{Q}$.
Subject to conditions (\ref{weakness}), calculation of the
required moments for the case of extended modes yields
\begin{equation}\label{Gv_moments}
  \Big<\left[G^{(V)}_{\bm{\mu}}(x,x')\right]^{p}\Big>
  =\left(\frac{-\mathrm{i}}{2\kappa_{\bm{\mu}}}\right)^{p}
  \exp\Bigg[  \mathrm{i}p  \kappa_{\bm{\mu}}|x-x'|-\frac{p}{2}
  \left(\frac{p}{L_f^{(V)}(\bm{\mu})}+
  \frac{1}{L_b^{(V)}(\bm{\mu})}\right)|x-x'|\Bigg] \ .
\end{equation}
Here, $L_{f,b}^{(V)}(\bm{\mu})$ are the forward ($f$) and backward
($b$) scattering lengths of the mode $\bm{\mu}$ which are
associated with the~prime intra-mode potential $V_{\bm{\mu}}(x)$,
\begin{equation}\label{ext_length}
  L_f^{(V)}(\bm{\mu})=\frac{4S}{9\mathcal{Q}}
  \left(2\kappa_{\bm{\mu}}\right)^2 \ ,\hspace{1.5cm}
  L_b^{(V)}(\bm{\mu})=\frac{4S}{9\mathcal{Q}}
  \frac{\left(2\kappa_{\bm{\mu}}\right)^2}%
  {\widetilde{\mathcal{W}}(\kappa_{\bm{\mu}})} \ ,
\end{equation}
$\widetilde{\mathcal{W}}(\kappa_{\bm{\mu}})$ is the Fourier
transform of $\mathcal{W}(x)$. As far as the evanescent modes are
concerned, at weak scattering (WS) the potential $V_{\bm{\mu}}(x)$
can be omitted from~(\ref{trialGV}), thus allowing one to take
advantage of the unperturbed solution,
\begin{equation}\label{evan}
  G^{(V)}_{\bm{\mu}}(x,x')=-\frac{1}{2|\kappa_{\bm{\mu}|}}
  \exp\big(-|\kappa_{\bm{\mu}}||x-x'|\big) \ .
\end{equation}

The functional structure of $T$-matrix (\ref{T-oper}) and,
consequently, of equation (\ref{GDIAG-FIN}) is substantially
simplified with the proviso (\ref{weakness}). Direct estimation,
with the use of (\ref{kernR}) and (\ref{Gv_moments}), of the
operator $\hat{\mathsf R}$ norm in the space of reference
functions $\{\exp(\mathrm{i}\kappa_{\bm{\mu}}x)\}$ is written as
\begin{equation}\label{R-norm}
  \|\hat{\mathsf R}\|^2\sim\frac{1}{\kappa_{\bm{\mu}}\ell}\ .
\end{equation}
This enables us to replace the exact operator $\hat{\mathsf K}$,
specified by equation (\ref{K_numu}), with its approximate value
$\hat{\mathsf K}\approx\hat{\mathsf R}$. As a result, the
potential (\ref{T-oper}) assumes the form
\begin{equation}\label{T-approx}
  \hat{\mathcal T}_{\bm{\mu}}=\bm{P}_{\bm{\mu}}\hat{\mathcal U}
  \hat{\mathcal G}^{(V)}\hat{\mathcal U}\bm{P}_{\bm{\mu}} \ ,
\end{equation}
where the operator $\hat{\mathcal G}^{(V)}$ is specified on
${\mathsf{\overline M}_{\bm{\mu}}}$ by matrix elements
\begin{equation}\label{GV-matr}
  \ket{x,\bm{\nu}}\hat{\mathcal G}^{(V)}\bra{x',\bm{\nu}'}=
  G^{(V)}_{\bm{\nu}}(x,x')\delta_{\bm{\nu}\bm{\nu}'}\ .
\end{equation}

The analogous substitution of the operator $\hat{\mathsf K}$
approximate matrix elements into Eq.~(\ref{nD_cond}) allows one to
conclude that at weak scattering the non-diagonal part of the
conductance is parametrically small as compared to its diagonal
counterpart. This is also confirmed by numerical calculation of
both of the conductance terms. Considering this fact, we further
restrict ourselves to the analysis of the term (\ref{D_cond})
only, assuming that $g(L)\approx g^{(d)}(L)$.

\section{Analysis of the mode states spectrum}

Unlike the original potential $V(\mathbf{r})$ and, accordingly,
its mode matrix elements in equation (\ref{mode_eqn}) the
effective potential $\hat{\mathcal T}_{\bm{\mu}}$ has a non-zero
mean value. In what follows, to apply the perturbation theory in
this potential one has to separate its averaged and fluctuating
parts, $\langle\hat{\mathcal T}_{\bm{\mu}}\rangle$ and
$\Delta\hat{\mathcal T}_{\bm{\mu}}=\hat{\mathcal
T}_{\bm{\mu}}-\langle\hat{\mathcal T}_{\bm{\mu}}\rangle$. As a
basic approximation for mode propagator
$G_{\bm{\mu}\bm{\mu}}(x,x')$ consider the Green function of the
equation
\begin{equation}\label{G(0)}
  \left[\frac{\partial^2}{\partial x^2}+
  \kappa^2_{\bm{\mu}}+{\rm i}0-
  \langle\hat{\mathcal T}_{\bm{\mu}}\rangle\right]
  G^{(0)}_{\bm{\mu}\bm{\mu}}(x,x')=\delta(x-x') \ ,
\end{equation}
which differs from (\ref{GDIAG-FIN}) by the lack of the
fluctuating potentials. On defining the operator
$\langle\hat{\mathcal T}_{\bm{\mu}}\rangle$ action onto the
function $G^{(0)}_{\bm{\mu}\bm{\mu}}(x,x')$ it is important to
notice that there is no correlation between inter- and intra-mode
scattering in the waveguide with hard side boundaries, viz. the
equality holds
\begin{equation}\label{inter-intra}
  \left<U_{\bm{\mu}\bm{\nu}}(x)V_{\bm{\nu}}(x')\right>=0\ .
\end{equation}
Owing to this, the potential (\ref{T-approx}), subject to
configurational averaging, is transformed from a non-local
operator to a~multiplicative constant. It was shown
in~\cite{bib:Tar00} that its effect on the mode energy
$\kappa^2_{\bm{\mu}}$ is reduced to occurrence of the mode
self-energy $\Sigma(\kappa_{\bm{\mu}})=\Delta
\kappa_{\bm{\mu}}^2+\mathrm{i}/\tau_{\bm{\mu}}^{(\varphi)}$,
\begin{equation}\label{self-en}
  \left(\langle\hat{\mathcal T}_{\bm{\mu}}\rangle
  \hat{G}^{(0)}_{\bm{\mu}\bm{\mu}}\right)(x,x')=
  -\Sigma(\kappa_{\bm{\mu}})G^{(0)}_{\bm{\mu}\bm{\mu}}(x,x')\ .
\end{equation}
By reproducing the calculation procedure given in \cite{bib:Tar00}
with reference to the system discussed in this paper one can
obtain
\begin{subequations}\label{renorm_spec}
\begin{eqnarray}
\label{ren_en}
  \Delta \kappa_{\bm{\mu}}^2 &=&
  \frac{\mathcal{Q}}{S}\sum_{\bm{\nu}\neq\bm{\mu}}\mathcal{P}
  \int_{-\infty}^{\infty}\frac{\mathrm{d}q}{2\pi} \;
  \frac{\widetilde{\mathcal{W}}(q+\kappa_{\bm{\mu}})}{q^2-\kappa_{\bm{\nu}}^2}
  \ , \\
\label{dephase}
  \frac{1}{\tau_{\bm{\mu}}^{(\varphi)}} &=&
  \frac{\mathcal{Q}}{4S}
  \overline{\sum_{\bm{\nu}\neq\bm{\mu}}} \frac{1}{\kappa_{\bm{\nu}}}
  \left[\widetilde{\mathcal{W}}(\kappa_{\bm{\mu}}-\kappa_{\bm{\nu}})+
  \widetilde{\mathcal{W}}(\kappa_{\bm{\mu}}+\kappa_{\bm{\nu}})\right]
  \ .
\end{eqnarray}
\end{subequations}
In (\ref{ren_en}), the symbol $\mathcal{P}$ stands for the
integral principal value.

The absolute value of the self energy (\ref{renorm_spec}) prove to
be not rather sensitive to the number of open channels. At any
$N_c>1$ the estimate $\Delta \kappa_{\bm{\mu}}^2\sim
1/\tau_{\bm{\mu}}^{(\varphi)}\sim k_F/\ell$ holds, which nearly
for all modes allows disregarding the mode energy renormalization
given by the term (\ref{ren_en}). At the same time, the mode level
uncertainty, (\ref{dephase}), is of crucial importance for further
analysis of the electron dynamics. The level width, apart from
being finite in magnitude, is quick to saturate with a growth in
the number of open channels. Specifically, within the model of
point-like scatterers the asymptotic of the term (\ref{dephase})
at $N_c\gg 1$ reads
\begin{equation}\label{deph_sat}
  \frac{1}{\tau_{\bm{\mu}}^{(\varphi)}}\approx\frac{\mathcal{Q}k_F}{4\pi}=
  \frac{k_F}{\ell}\ .
\end{equation}

Note particularly that in (\ref{dephase}), as opposed to
(\ref{ren_en}), the summation is carried out over \emph{extended}
modes only of the quantum waveguide, the mode $\bm{\mu}$ itself
being excluded as an intermediate state. In the case of a
single-channel conductor, where solely the lowest mode
$\bm{\mu}_1=(1,1)$ is non-local, the self energy is free of the
term (\ref{dephase}).

The imaginary part of self energy (\ref{renorm_spec}) can be
interpreted as an effect of the coherent mode state dephasing.
From expression (\ref{dephase}) structure it follows that for any
given mode $\bm{\mu}$ the re-entrant electron scattering through
the non-local intermediate modes, with the proviso that the latter
are present in the system at hand, serves as a cause of dephasing.
This suggests the interpretation that for any current-carrying
mode in the conductor all other \emph{extended} modes, except the
mode $\bm{\mu}$ itself, can be thought of as a peculiar
``dephasing environment''. Interaction of modes with this
environment is realized via the \emph{inter-mode scattering} from
the potential $V(\mathbf{r})$. Although the scattering from static
disorder is certainly elastic in terms of a one-electron energy
(at $T=0$ the latter remains Fermian), the \emph{many-particle}
mode states are specified by distinct (longitudinal) energies.
This suggests considering the virtual inter-mode transitions,
``hidden'' in $T$-matrices (\ref{T-oper}) and (\ref{T-approx}), as
being effectively inelastic, thereby making it possible to adhere
to the conventional point of view which does associate the
dephasing of quantum states solely with inelastic processes.

Note that the inter-mode scattering via solely strongly localized
evanescent modes (single-mode conductor) does not result in
dephasing the mode states. The dephasing effect is noticeable
provided electrons are scattered through essentially non-local
extended modes, which is certainly possible if there are at least
two such modes in the conductor. This leads us to conclude that
the preservation of quantum states \emph{spatial coherence} is
every bit as important for \emph{interferential} Anderson
localization as the time coherence.

The effect of fluctuating potentials, $V_{\bm{\mu}}(x)$ and
$\Delta\hat{\mathcal T}_{\bm{\mu}}$, can be analyzed by evaluating
the corresponding Born scattering rates, $1/\tau_{\bm{\mu}}^{(V)}$
and $1/\tau_{\bm{\mu}}^{(\mathcal T)}$. Taking into account the
operator (\ref{T-approx}) structure and the estimate
(\ref{deph_sat}) one can readily obtain
\begin{subequations}\label{estim-rates}
\begin{eqnarray}
  \label{estim-rates1}
  \frac{\tau_{\bm{\mu}}^{(\varphi)}}{\tau_{\bm{\mu}}^{(V)}}
  &\sim& \left[k_Fr_cN_c\cos^2\vartheta_{\bm{\mu}}\right]^{-1} \ , \\
  \label{estim-rates2}
  \frac{\tau_{\bm{\mu}}^{(\varphi)}}{\tau_{\bm{\mu}}^{(\mathcal{T})}}
  &\sim& \frac{\tau_{\bm{\mu}}^{(\varphi)}}{\tau_{\bm{\mu}}^{(V)}}
  \times\left\{
  \begin{array}{ccc}
    L/\ell &,&\qquad L<\ell \\
    N_{loc}^{(s)} &,&\qquad L>\ell
  \end{array}\right. \ .
\end{eqnarray}
\end{subequations}
Here, $\vartheta_{\bm{\mu}}$ is the mode $\bm{\mu}$ angle of slide
($\cos\vartheta_{\bm{\mu}}=|\kappa_{\bm{\mu}}|/k_F$),
$N_{loc}^{(s)}\leq N_c$ is the number of trial mode states
corresponding to equation (\ref{trialGV}), whose localization
lengths, $4L_b^{(V)}(\bm{\mu})$, do not exceed the conductor
length~$L$.

Since in real materials the condition $k_Fr_c\gtrsim 1$ normally
holds, it is easy to make sure that the potential
$V_{\bm{\mu}}(x)$ has a negligible effect on the electron dynamics
in multi-mode systems. The same applies to the potential
$\Delta\hat{\mathcal T}_{\bm{\mu}}$ if one assumes the conductor
length $L\ll N_c\ell$. The comprehensive analysis performed in
\cite{bib:Tar00} has revealed that the scattering on this
potential (substantially non-local in the case of $N_c>1$) does
not either affect the conductance significantly if $L\gg N_c\ell$.
This seems to be quite natural if one bears in mind that the
potential $\Delta\hat{\mathcal T}_{\bm{\mu}}$ brings about the
\emph{re-entrant} scattering. From the standpoint of perturbation
theory this implies that in scattering from the potential
$\Delta\hat{\mathcal T}_{\bm{\mu}}$ the inter-mode potentials
$U_{\bm{\mu}\bm{\nu}}(x)$ contribute twice as much as they do when
self energy (\ref{renorm_spec}) is obtained.

Turning back to the prime intra-mode potential $V_{\bm{\mu}}(x)$,
it should be stressed that, subject to the condition
(\ref{weakness}), the mode states spatial coherence is slightly
violated by scattering on this potential. In multi-mode conductors
($N_c\gg 1$) this scattering leads to weak localization
corrections to the conductance, which are not dealt with in this
paper. Clearly, the role of those corrections increases with a
decreasing number of conducting channels, but they do not
considerably affect the result obtained within the kinetic
approach even at $N_c\sim 1$. The potential $V_{\bm{\mu}}(x)$ is
determining for the electron spectrum in \emph{one-dimensional}
conductors only, which will be discussed in more details in the
next section.

\section{The conductance dependence on the conductor con\-fi\-guration}

The mode state spectrum is governed, along with the electron
energy, by the confinement potential configuration, i.e. by the
conductor geometry, according to the model being considered. While
the bulk wire conductance changes, with a variation of the
conductor shape, in compliance with the conventional Ohm's law, in
the case where at least one of the conductor dimensions is
comparable to microscopic length scales pertinent to the system at
hand the quantization becomes extremely important. Consider some
limiting cases where the dimensional quantization affects the
electron dynamics quite differently.

\subsection{Multi-mode conductors}

If the confinement potential and the electron energy are such that
there is more than one conducting channel in the quantum
waveguide, the exact Green function $G_{\bm{\mu}\bm{\mu}}(x,x')$
can be shown, including (\ref{estim-rates}), to be
well-approximated in the range $L\ll N_c\ell$ by the function
$G^{(0)}_{\bm{\mu}\bm{\mu}}(x,x')$. In the region $L>N_c\ell$ the
replacement of the exact propagator in (\ref{D_cond}) by its
approximate form from equation (\ref{G(0)}) is yet justified in
view of statistical averaging over the disorder. The solution of
(\ref{G(0)}) which meets the radiation conditions at open ends of
the conductor has the form
\begin{equation}\label{Gmm0}
  G_{\bm{\mu}\bm{\mu}}^{(0)}(x,x')=\frac{1}{2\mathrm{i}\kappa_{\bm{\mu}}}
  \exp\Big\{\big[\mathrm{i}\kappa_{\bm{\mu}}-1/l_{\bm{\mu}}^{(\varphi)}\big]
  |x-x'|\Big\} \ .
\end{equation}
Here,
$l_{\bm{\mu}}^{(\varphi)}=2\kappa_{\bm{\mu}}\tau_{\bm{\mu}}^{(\varphi)}$
is the mode $\bm{\mu}$ extinction length (or, equivalently, its
dephasing length) associated with the incoherent inter-mode
scattering. Substitution of (\ref{Gmm0}) into (\ref{D_cond})
results in the average conductance expression as follows,
\begin{equation}\label{g(d)}
\big<g(L)\big>=
  \overline{\sum_{\bm{\mu}}}\frac{l_{\bm{\mu}}^{(\varphi)}}{L}
  \left[1-\frac{l_{\bm{\mu}}^{(\varphi)}}{L}
  \exp\left(-\frac{L}{l_{\bm{\mu}}^{(\varphi)}}\right)
  \sinh\frac{L}{l_{\bm{\mu}}^{(\varphi)}}\right] \ .
\end{equation}

\begin{figure}[h]
\centering \scalebox{.6}[.6]{\includegraphics{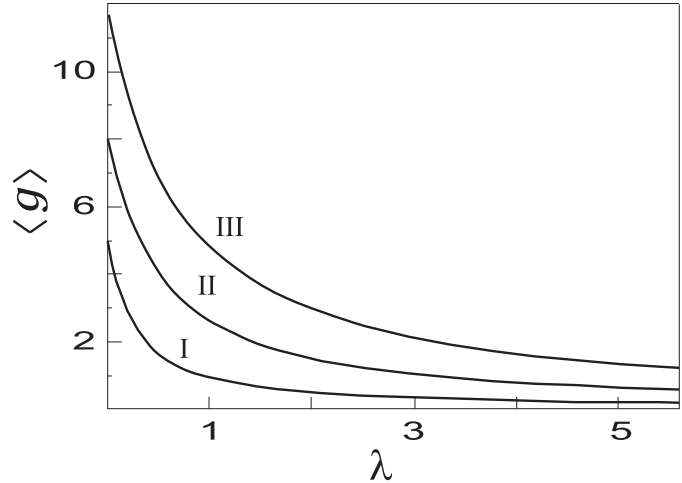}}
\caption{The conductance (\ref{g(d)}) as a function of the
dimensionless length $\lambda=L/\ell$ for conductors with
different number of open channels: I --- $N_c=5$, II
--- $N_c=8$, III --- $N_c=12$.} \label{Fig2}
\end{figure}

If the number of channels is large, $N_c\gg 1$, the replacement of
the sum (\ref{g(d)}) by the integral results in simple limiting
formulas valid in the regions corresponding to classically
ballistic ($L\ll\ell$) and diffusive ($L\gg\ell$) electron
transport, namely
\begin{subequations}\label{cond_asymp}
\begin{eqnarray}
\label{ball}
  &&\big<g(L)\big>\approx N_c\qquad\ ,\hspace{1cm}L\ll\ell\ , \\
\label{diff}
  &&\big<g(L)\big>\approx \frac{4}{3}\frac{N_c\ell}{L}\quad,\hspace{1cm}L\gg\ell\ .
\end{eqnarray}
\end{subequations}
In the ballistic limit (\ref{ball}), the conductance as a function
of the electron energy or the confinement potential has a
staircase-like structure with the step height being equal exactly
to the conductance quantum $G_0=e^2/\pi\hbar$ (recall that for a
massive tree-dimensional wire the equality $N_c=[k_F^2S/4\pi]$
holds, where $[...]$ stands for the integer part of the number
enclosed). As the electron motion changes from ballistic to
diffusive regime, the conductance approaches asymptotically the
classical value (\ref{diff}) well known from the kinetic theory.
Staircase structure of the conductance is formally kept safe, but
the step height is decreased in proportion to the ratio $\ell/L$.
The conductance dependence on the quantum waveguide length is
displayed in Fig.~\ref{Fig2}. The curves correspond to different
numbers of open channels. Nevertheless, all of them show the same
``ohmic'' behaviour at lengths obeying $L/\ell>1$.

\subsection{Anderson localization in a single-mode conductor}

If the electron system parameters permit one open channel only,
all other modes of the quantum waveguide are inhomogeneous,
strongly localized along the $x$ coordinate (so-called evanescent
modes). In this case the potential $\hat{\mathcal T}_{\bm{\mu}}$,
as well as $V_{\bm{\mu}}(x)$, is real-valued and local. Therefore,
the perturbation theory in the form used above ceases to be
practical since the weak scattering, including the inter-mode one,
does not substantially violate the coherence of each extended
mode. Calculation of the conductance in this case calls for a
method which would take into account the interference of multiply
scattered quantum waves, e.g., the methods used in
\cite{bib:B73,bib:AR78} to obtain the conductivity of 1D
disordered conductors.

In \cite{bib:MT98,bib:MT01}, with the use of the resonance weak
scattering method equivalent to those of \cite{bib:B73,bib:AR78},
the general expression for the entire set of statistical moments
of the single-mode wire conductance was obtained in the case of
the disorder produced by side boundary roughness of the conductor
instead of its bulk inhomogeneities. Technically, the
rough-boundary problem is identical to that considered in this
paper, so application of the method \cite{bib:MT98,bib:MT01} to
a~single-mode bulk-disordered conductor yields
\begin{eqnarray}\label{g-moments}
  \big< g^n(L)\big> &= & \frac{4}{\sqrt\pi}
  \left[\frac{L_b^{(V)}(\bm{\mu}_1)}{L}\right]^{3/2}
  \exp{\left[-\frac{L}{4L_b^{(V)}(\bm{\mu}_1)}\right]}
  \int_0^\infty\frac{zdz}{\cosh^{2n-1}z}
  \exp\left[-z^2\frac{L_b^{(V)}(\bm{\mu}_1)}{L}\right] \int_0^z dy\,
  \cosh^{2(n-1)}y \, ,\qquad
\\[15pt]
 n &=& 0,\pm 1,\pm2,\ldots  \ . \nonumber
\end{eqnarray}
It can be concluded from (\ref{g-moments}) that in the one-channel
case two regimes only of the electron transport can be
distinguished, viz. ``ballistic'' and ``localized'', the
corresponding limiting expressions for the average conductance
being
\begin{equation}\label{cases}
  \langle g(L) \rangle \approx
\begin{cases}
  1-4L/\xi_1\ , &\qquad L/\xi_1\ll 1 \\
  (\pi^{5/2}/16)\left(\xi_1/L\right)^{3/2}
  \exp\left(-L/\xi_1\right)\ ,&\qquad L/\xi_1\gg 1
\end{cases}\ .
\end{equation}
Here, $\xi_1=4L_b^{(V)}(\bm{\mu_1})$ is the harmonics $\bm{\mu}_1$
\emph{one-dimensional} localization length associated with the
electron backscattering on the prime intra-mode potential
$V_{\bm{\mu}_1}(x)$.

\subsection{Metal-insulator transition as a quantum phase transition}

The conductance (\ref{cases}) evidently exhibits the localized
character of the electron transport in a single mode quantum
waveguide, in accordance with the well-known results of 1D
disordered system spectral analysis \cite{bib:LGP82}. This fact in
itself suggests that it is in principle possible for the finite
electron system to be transferred from conducting to insulating
state subject to the geometrical factors only, the disorder being
kept constant. The one-dimensional Anderson-type localization is
known to be universal in linear systems, in that all the electron
states in 1D random potential are localized in infinite-length
systems, irrespective of the electron energy. At the same time,
this localization, in a sense, can be considered to be weak. On
lowering the disorder level the length $\xi_1$ increases
infinitely, so that in relatively perfect conductors, their length
being even extremely large, the collective motion of the electrons
can actually remain nearly ballistic.

The approach suggested in this paper provides an explanation for
the observed MIT even at moderate, viz., mesoscopic, sample
lengths, the disorder being arbitrary enough. When decreasing the
cross-section area, the conductor should ultimately turn to a
below-cut-off waveguide regime where all of the modes become
evanescent, each being localized at a scale of the wavelength
$|\kappa_{\bm{\mu}}|^{-1}$ which, in the context of calculation
technique being used, is considered as microscopic. In such a
``dimensionally localized'' regime, the conductance falls sharply
with respect to its value (\ref{cases}) in the ``marginal''
one-mode state, thus allowing for being regarded as equal to zero
with parametric accuracy.

It is essential that the mode spectrum of the conductor depicted
in Fig.~\ref{Fig1}b can be varied by changing just one of its
transverse dimensions, the other being kept constant. From
(\ref{kappa_munu}) it can be seen that even at large enough width
$W$ the quantum waveguide can be carried over to the below-cut-off
regime subject to the decrease of its thickness $H$ only. In real
planar systems this is achieved by either increasing the depletion
voltage (see Fig.~\ref{Fig1}a) or by acting upon the heterocontact
region capacitively.

\begin{figure}[h]
\hspace{-.2cm} \centering
\scalebox{.85}[.85]{\includegraphics{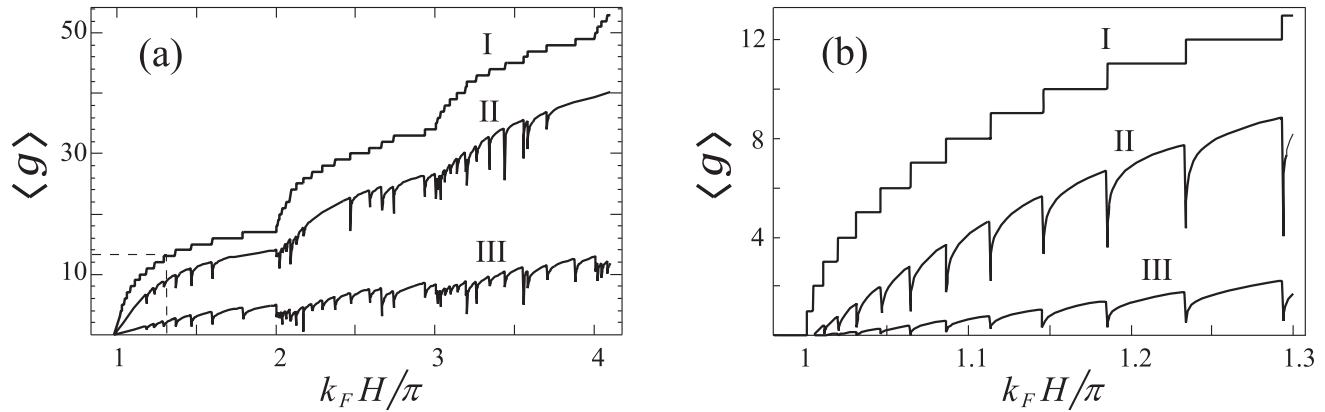}} \caption{The
dimensionless conductance vs the quantum waveguide thickness, at
fixed width ($k_FW/\pi=~20.5$) and different values of the
diffusion parameter $\lambda=L/\ell$. The curve I corresponds to
$\lambda=0$, II to $\lambda=0,5$, III to $\lambda=5$.}
\label{Fig3}
\end{figure}

In Fig.~\ref{Fig3}, the numerical data are presented showing the
conductance (\ref{g(d)}) dependence on the conductor thickness,
the width being fixed. The curve I conforms to the ballistic limit
$\ell/L\to\infty$, whereas the curves II and III to the finite
values of this ratio. The ballistic conductance is ideally
quantized, each step being equal to the quantum $G_0$. The
peculiar slow modulation of the curves is due to the waveguide
model used, Fig.~\ref{Fig1}b, whose spectrum requires the opening
or closing of the conducting channels to occur non-equidistantly
in the value of $H$.

As the disorder increases (the curves II and III), the conductance
steps become lower, their contours being smoothed out. In the
vicinity of the channel opening (closing) points significant dips
in the conductance should be observed. The shape of the dips can
be clearly seen in Fig.~\ref{Fig3}b, where the region separated in
Fig.~\ref{Fig3}a is shown at the expanded scale. When approaching
the channel closing point from the large $H$ side, the conductance
slowly decreases. This takes place due to the increase in the
density of states of the slow marginal mode $\bm{\mu}_m$ and,
consequently, due to the electron transfer to this mode from more
``fast'' open channels. The dephasing rates of the latter,
Eq.~(\ref{dephase}), have square-root singularities at the
critical points, which results in the destructive reduction of the
mode propagator (\ref{Gmm0}) when approaching the point
corresponding to $\kappa_{\bm{\mu}_m}=0$.

The analogous dips were found in the waveguide system optical
conductance calculated numerically using the Landauer approach
\cite{bib:GSGC96}. However, in \cite{bib:GSGC96} the dips were
relatively symmetric with respect to the points of extended mode
disappearance, whereas in Fig.~\ref{Fig3} they appear to be
pronouncedly asymmetric in form. The skewness is explained by the
fact that in deriving the expression (\ref{g(d)}) in the WS
approximation we have disregarded the evanescent mode contribution
to the conductance, thus omitting the tunnel part of the
conductance which is governed by those modes. This is not quite
justified for marginal modes since the WS condition for them is
violated at the critical point, so that just after the mode
closing its propagator does not equal exactly to (\ref{evan}).

The graphs in Fig.~\ref{Fig3} display a succession of quantum
phase transitions which take place in the electron system if one
changes the confinement potential. At the critical points the
conductance varies stepwise, the marginal mode wavelength serving
as a correlation length in the electron system.

The leftmost phase transition, clearly seen in Fig.~\ref{Fig3}b,
can be interpreted as the electron system transition from metallic
to dielectric state. In the metallic phase, straight to the right
of the transition point, the conductance in the ideal ballistic
situation equals exactly to the quantum $G_0$. This is in a good
agreement with the resistance values observed alongside the
so-called separatrix, the conventional line separating the
ensembles of experimental curves showing the resistance
temperature dependence in dielectric and conducting phases of
two-dimensional systems~\cite{bib:AKS01}.

In most of the experimental works the 2D systems spectral
classification is performed on the bases of the temperature and
the magnetic field dependence of their resistance. The detailed
analysis of the magnetic-field-induced effects is beyond the scope
of this paper. As far as the resistance temperature dependence is
concerned, some qualitative conclusions can be made in accordance
with the above described peculiarities of the quantum transport in
planar systems.

It should be noted that the transition from the metallic-type
conductance (\ref{g(d)}), (\ref{cond_asymp}) to its small value in
the localized ($0$-mode) phase proceeds inevitably through the
one-mode state of the electron system, the latter behaving like
the effectively one-dimensional in spite of the macroscopic width.
It has already been predicted, in \cite{bib:AR78} for the case of
$T\tau>>1$ and in \cite{bib:T92} for $T\tau<<1$, that a 1D system
conductivity should exhibit non-monotonic dependence on $T$. And
in fact, the actual 2D system resistance measured on the metallic
side close to the separatrix tends to change non-monotonically
with temperature \cite{bib:AKS01}.

The weak temperature dependence of the separatrix itself can also
be explained if one takes into account that the mode wavelength of
the last-remaining extended mode grows infinitely as far as one
approaches its closing point. Since this length prove to exceed
ultimately the wavelength of thermal phonons, the interaction
between the marginal-mode electrons, whose density of states grows
proportionally to the mode wavelength, turns out to be
ineffective.

In conclusion, deep in the dielectric phase, where all the
electron modes become evanescent, that is, strongly localized in
the direction of current, it is natural to anticipate the
resistance temperature dependence predicted by the percolation
theory \cite{bib:SE79}. The dependence similar to this type is
widely observed in real two-dimensional systems far in the
dielectric side from the separatrix \cite{bib:AKS01,bib:MKBF95}.

\section{Concluding remarks}

The objective of the present paper was to elaborate a one-particle
field model of a 2D electron system transition from the dielectric
state suggested by the scaling theory of localization to a
metallic phase widely observed in experiment. The essence of the
proposed approach lies in the fact that electrons (or holes),
being experimentally regarded as evidently two-dimensional, should
be really believed to be propagating in a virtual
three-dimensional quantum waveguide formed by the confinement
potential. Within the framework of this approach, the quantum
nature of the electrons can be fully taken into account.

Note that the technique of a multi-dimensional problem reduction
to a set of strictly one-dimensional problems, which is the key
point of the suggested analytical procedure, is also applicable to
those systems which are initially considered as strictly
two-dimensional ones \cite{bib:Tar99,bib:Tar00}. However, the
possibility for such systems to change from the metallic to
dielectric state can hardly be noticed within the framework of
this technique. The point is that even a single-mode state of a 2D
electron system, let alone its $0$-mode regime, is generally
associated not with a~macroscopic conductor but rather with an
extremely narrow strip-like quantum wire.

In spite of such a perception, a macroscopic two-dimensional
quantum waveguide, being actually considered as a flattened
three-dimensional one, can be easily reduced to the single-mode
and then even to the below-cut-off ($0$-mode) state. Thereupon,
the question is bound to arise which electron systems can be
reasonably attributed to a class of two-dimensional systems and
which cannot be. Moreover, is there, in fact, the essential
difference, from the transport properties standpoint, between any
non-one-dimensional electron systems and three-dimensional ones?

It is difficult to ascertain the objective criteria for
distinguishing between 2D and 3D electron systems based solely on
the results given in this paper. The point is that only the
diffusion-type conductance typical for non-single-mode real
disordered systems and the ``localized'' conductance pertinent to
one-mode conducting as well as $0$-mode (dielectric) systems seem
to be fundamentally different. Additionally, if one considers that
2D and 3D transport problems in the mode representation are
similarly reduced to one-dimensional ones, the conclusion suggests
itself that it might be logical to classify the non-ballistic
systems of Fermian type in terms of the mode content instead of
their formal geometrical structure. From this standpoint, two
kinds of systems seem to be, in actual fact, fundamentally
different, viz. systems possessing more than one extended mode
(both two- and three-dimensional), wherein the diffusive
quasi-particle transport is realized, and those which can be
conditionally referred to as the ``localized'' systems. The latter
class includes one-mode systems subject to Anderson-type
localization (weak or strong, depending on the disorder strength)
and $0$-mode systems in which the localization is merely related
to the \emph{dimensional quantization} of the electron spectrum
rather than to the disorder and/or Coulomb interaction of
carriers.


\end{document}